\documentclass[12pt,preprint]{aastex}
\begin{document}

\title{\bf The Coronal Abundance Anomalies of M Dwarfs}

\author{Brian E. Wood\altaffilmark{1}, J. Martin Laming\altaffilmark{1},
  Margarita Karovska\altaffilmark{2}}


\altaffiltext{1}{Naval Research Laboratory, Space Science Division,
  Washington, DC 20375, USA; brian.wood@nrl.navy.mil}
\altaffiltext{2}{Smithsonian Astrophysical Observatory, 60 Garden St.,
  Cambridge, MA 02138, USA}

\begin{abstract}

     We analyze {\em Chandra} X-ray spectra of the M0~V+M0~V binary
GJ~338.  As quantified by X-ray surface flux, these are the most
inactive M dwarfs ever observed with X-ray grating spectroscopy.  We
focus on measuring coronal abundances, in particular searching for
evidence of abundance anomalies related to First Ionization Potential
(FIP).  In the solar corona and wind, low FIP elements are
overabundant, which is the so-called ``FIP effect.''  For other stars,
particularly very active ones, an ``inverse FIP effect'' is often
observed, with low FIP elements being underabundant.  For both members
of the GJ~338 binary, we find evidence for a modest inverse FIP
effect, consistent with expectations from a previously reported
correlation between spectral type and FIP bias.  This amounts to
strong evidence that {\em all} M dwarfs should exhibit the inverse FIP
effect phenomenon, not just the active ones.  We take the first step
towards modeling the inverse FIP phenomenon in M dwarfs, building on
past work that has demonstrated that MHD waves coursing through
coronal loops can lead to a ponderomotive force that fractionates
elements in a manner consistent with the FIP effect.  We demonstrate
that in certain circumstances this model can also lead to an inverse
FIP effect, pointing the way to more detailed modeling of M dwarf
coronal abundances in the future.

\end{abstract}


\section{INTRODUCTION}

     The solar corona and solar wind exhibit a curious abundance
pattern in which the abundance of an element relative to its abundance
in the photosphere is dependent on its first ionization potential
(FIP).  Elements with low FIP (Fe, Mg, Si, etc.) are found to have
coronal abundances that are enhanced relative to elements with high
FIP (C, N, O, Ne, etc.), by about a factor of four on average
\citep{rvs95,uf00}.  This abundance anomaly
has been called the ``FIP effect,'' which at least on the Sun is
known to be due to enhancement of low-FIP elements relative to
photospheric abundances, as opposed to depletion of high-FIP elements.
This phenomenon has been the
subject of many theoretical studies. Some of these attribute it to the
effects of Alfv\'{e}n waves passing through the solar atmosphere,
which may also be involved in coronal heating \citep{jml04,jml09,jml12}.
In addition to its presence on the Sun, evidence for a FIP
effect has also been found for some stars of low to moderate activity
\citep{jml96,jjd97,jml99}.
However, on very active stars the FIP effect tends to be either
absent, or sometimes an inverse FIP effect is observed, where low-FIP
elements have coronal abundances that are {\em depleted} relative to the
high-FIP elements \citep{ma01,ma03,acb01,mg01,dph01,dph03,jsf03,jsf09,bb05}.
So far this behavior is not
captured in any theoretical model. However, the generalization of the
\citet{jml09,jml12} models to include other MHD wave modes is expected
to yield an inverse FIP effect in the appropriate conditions, within
the same framework as that for the FIP effect.

     The impression these results provide is of coronal abundances
being primarily dependent on stellar activity, with FIP effect
changing to inverse FIP as activity increases.  However, if only main
sequence stars are considered, and if extremely active stars with
X-ray luminosities of $\log{L_X}>29$ (in ergs~s$^{-1}$) are excluded,
then all activity dependence disappears, and one instead sees a strong
dependence on spectral type \citep[][hereafter WL10]{bew10}.  In this
correlation, early G stars all have a solar-like FIP effect, which
decreases towards late G and early K stars, reaching no FIP effect at
all at a spectral type of K5~V.  Later than K5~V, inverse FIP effects
are observed, with the magnitude of the effect increasing with
spectral type.  We will refer to this as the
``FIP-Bias/Spectral-Type'' (FBST) relation.  This suggests that for
the vast majority of cool stars in the Galaxy, coronal abundances are
determined entirely by spectral type, and not stellar age or activity.
This illustrates the problems inherent in the dominance of very active
stars in the archives of X-ray spectra.  The stars that are brightest
and most easily observed by X-ray observatories are not necessarily
representative of the Galactic stellar population as a whole, and
therefore may provide a misleading picture of the nature of X-ray
emitting stellar coronae.

     The FBST relation of WL10 has an important implication for M
dwarfs.  It implies that {\em all} M dwarf coronae should possess an
inverse FIP effect.  Relatively inactive stars (with $\log{L_X}<29$)
that follow the FBST relation should have it, and if there is an
activity dependence of FIP bias for the most active M dwarfs,
experience suggests that this would only act to increase the inverse
FIP effect even further.

     One problem with verifying this prediction for M dwarfs is that
only relatively active M dwarfs can be observed.  This is a problem to
some degree for all stars, of course, but the problem is magnified for
M dwarfs because of their small size.  Consider Proxima~Centauri
(M5.5~Ve), which is by any measure the most inactive M dwarf
previously observed spectroscopically in X-rays.  With an X-ray
luminosity of $\log L_X=27.22$ \citep{js04} it is only by
virtue of its incomparable close proximity ($d=1.3$~pc) that it can
be observed spectroscopically at all.  However, although Proxima~Cen
seems quite inactive as quantified by $L_X$, this is not the case if
one quantifies activity by surface X-ray flux, $F_X$.  With a radius
of only 0.15 R$_{\odot}$ \citep{ds03}, Proxima~Cen's
low X-ray luminosity translates to a relatively high X-ray surface
flux of $\log F_X=6.08$ (in ergs~cm$^{-2}$~s$^{-1}$).  This $F_X$ is
higher than that of most of the stars that WL10 used to define the
FBST correlation, though still lower than all the
other M dwarfs.  Proxima~Cen is the most inactive M dwarf known
to possess an inverse FIP effect, but its relatively high $F_X$ value
leaves open the possibility that at sufficiently low $F_X$, the
inverse FIP effects of the M dwarfs will drift towards a more
solar-like FIP effect.

     Our primary goal in this paper is to analyze new {\em Chandra}
spectra of the GJ~338 binary (M0~V+M0~V), consisting of two M dwarfs
with $F_X$ much lower than Proxima~Cen.  We aim to determine whether
the GJ~338 coronae possess the expected inverse FIP effect.  If it
does, this would further demonstrate that the inverse FIP phenomenon
is {\em not} solely or even primarily an effect of high activity.  It
would also provide crucial support for the universality of the
FBST relation for inactive and modestly active
main sequence stars, and provide evidence that {\em all} M dwarf
coronae exhibit the inverse FIP effect.

\section{X-RAY IMAGING OF THE GJ 338 SYSTEM}

     Our target, GJ~338, was chosen for its close proximity
($d=5.81$~pc) and modest activity level, with an X-ray flux just high
enough to be observable spectroscopically.  The binarity of the target
also makes it attractive, as it allows two relevant stellar spectra to
be taken in a single exposure.  The only previous X-ray observation to
resolve the binary was by the ROSAT HRI imager, which found X-ray
luminosities for GJ~338A and GJ~338B of $\log L_X=28.16$ and $\log
L_X=28.21$, respectively.  However, weaker X-ray
emission was observed during the ROSAT All-Sky Survey (RASS), which
found the binary to have a combined $\log L_X=27.85$ \citep{js04}.

     {\em Chandra} observed GJ~338 on 2010~December~29 for 96.5~ksec,
using the LETGS setup (LETG grating plus HRC-S detector).  An LETGS
observation consists of a zeroth-order image plus two identical
spectra of the target stretched out to either side of the image along
the dispersion direction.  The zeroth-order image is shown in Figure~1a,
showing the two members of the binary with roughly equal brightness.
The measured position angle and stellar separation for the binary is
$\theta=96.41^{\circ}$ and $\rho=17.13^{\prime\prime}$, respectively.
X-ray luminosities can be estimated directly from the image, which
implies that both stars are much fainter than expected from the ROSAT/HRI
luminosities quoted above.  From the spectra discussed below we
measure X-ray luminosities within $5-120$~\AA\ ($0.1-2.4$~keV) of
$\log L_X=27.64$ and $\log L_X=27.60$ for GJ~338A and GJ~338B,
respectively, values lower than expected from the HRI observation,
but consistent with the RASS luminosity quoted above.
In order to quantify $F_X$ for these stars, we use
\citet{asg74} to estimate radii of 0.59~R$_{\odot}$ and
0.58~R$_{\odot}$ for GJ~338A and GJ~338B, respectively, leading
to surface fluxes of $\log F_X=5.31$ and $\log F_X=5.29$, values about
a factor of 6 lower than $F_X$ for Proxima~Cen.

     The zeroth-order images are also useful for quantifying source
variability, and Figure~1b shows light curves for the two stars.  The
secondary shows no significant X-ray variability, but a couple short
flares are apparent on the primary.  Such X-ray flaring is quite
common for M dwarfs \citep[e.g.,][]{rao05,afk09}.

     The GJ~338A image in Figure~1a appears to be asymmetric,
especially compared with that of GJ~338B.  In particular, there is
excess emission NNW of GJ~338A.  However, we have determined this to
be caused by a recently identified instrumental artifact of the HRC-S
detector, which is worth bringing to the attention of other HRC-S
users.  In short, the region of the HRC-S detector on which GJ~338A
ended up located has been found to have unique problems with event
positions, confirmed by HRC-S observations of the {\em Chandra}
calibration target HZ~43 (B.\ Wargelin, private communication).
As part of its system of identifying the
centroids of charge clouds induced by X-ray photons, the HRC-S
detector uses charge amplifiers called ``taps'' \citep{jhc89,mj00}.
The dithering pattern used during the course of our LETGS observation
carried GJ~338B across taps numbered 98 and 99, while GJ~338A was
covered by taps 98--100.  It is tap number 100 that is the problem,
which therefore affected GJ~338A without affecting GJ~338B.  The
aimpoint for HRC-S has drifted with time, so it is only recently that
this problematic region of the detector has moved to such a
disadvantageous location.  Note that the spectra themselves do not
fall on this part of the detector, so they are unaffected.

\section{X-RAY SPECTROSCOPY OF THE GJ 338 SYSTEM}

     For GJ~338, obtaining spectra from the LETGS data requires a
certain amount of specialized data processing, in order to extract
separate spectra for each member of the binary.  We follow procedures
similar to ones used in past analyses of binary stars observed by
LETGS \citep{bew06}, and we refer the reader to those analyses
for details about this data processing.  The spectra are processed
using version 4.3 of the CIAO software.  Although LETGS provides very
broad spectral coverage from $5-175$~\AA, we find that we are only able to
detect the strongest emission lines from these stars, which are below
35~\AA.  Thus, the decreasing spatial resolution with wavelength
exhibited by LETGS spectra is not a concern, so to maximize
signal-to-noise for the detected lines we use a relatively narrow
extraction window of 16 pixels in the cross-dispersion direction.

     Figure~2 shows the resulting spectra, which have been rebinned by
a factor of 5 and smoothed slightly to better reveal the emission
lines in the noisy data.  There are only eight emission features that
can be discerned.  Table~1 lists counts and line fluxes measured for
these lines, some of which are actually blends of multiple lines.

     Our principle aim here is to measure the FIP bias present in the
coronae of the two GJ~338 stars.  Ideally, this requires a full
emission measure analysis of line fluxes, as described by WL10.
However, with only a limited number of lines to work with, such an
analysis is clearly impossible here.  We instead estimate a FIP bias
from only the strongest lines in the spectrum, those of Fe~XVII and
O~VIII.  Since Fe is a low-FIP element while O is a high-FIP element,
the Fe~XVII/O~VIII flux ratio provides a reasonable estimate of FIP
bias, particularly since both Fe~XVII and O~VIII are formed at about
the same temperature of $\log T=6.6$.

     Between $15-17$~\AA, coronal spectra are typically dominated by
five lines of Fe~XVII, at rest wavelengths of 15.015~\AA, 15.262~\AA,
16.778~\AA, 17.053~\AA, and 17.098~\AA\ \citep{at05,bew06,bew10}.  In
the smoothed spectra in Figure~2, these lines coalesce into two
blended Fe~XVII features.  In order to explore how Fe~XVII/O~VIII
varies with FIP bias, we use the sample of stars studied by WL10.  A
list of these stars is provided in Table~2, with the addition of a few
extra ones (particularly $\alpha$~Cen~AB), as described below.  For
the WL10 stars, we sum the fluxes of the Fe~XVII lines and divide the
sum by the line flux observed for the H-like O~VIII Lyman-$\alpha$
line at 18.97~\AA, yielding a final Fe~XVII/O~VIII ratio.  These
ratios incorporate spectral measurements and analyses by
\citet{jun03}, \citet{at05}, and \citet{cl08}.

     The FIP biases defined by WL10 ($F_{bias}$) are listed in Table~2
and plotted versus the Fe~XVII/O~VIII ratios in Figure~3.  As
described in detail by WL10, the $F_{bias}$ number represents an
attempt to quantify the FIP bias in a stellar corona in a single
number that considers abundance measurements of four high-FIP elements
(C, N, O, and Ne) and the best reference low-FIP element, Fe.  For
each of the four high-FIP elements we take the logarithmic abundance
ratio with Fe, minus the assumed photospheric ratio, i.e., $\log {\rm
[X/Fe]}-\log {\rm [X/Fe]}_*$.  The average of these four quantities is
$F_{bias}$.  No photospheric abundance measurements are available for
the M dwarfs in the sample, including GJ~338, so we simply have to
assume solar photospheric abundances apply.  \citet{ma09} is used as
the source of solar abundances, but for reasons described in detail in
WL10, we replace the Ne abundance with the one suggested by the
average coronal Ne/O ratio measured by \citet{jjd05}.  For the GK
dwarfs, stellar photospheric abundances from \citet{cap04} are used,
but these abundances are measured relative to solar ones, so even for
these stars the assumed reference solar abundances must be used to
compute $\log {\rm [X/Fe]}_*$.

     As quantified this way, $F_{bias}=0$ corresponds to coronal
abundance ratios being identical (on average) to photospheric ratios,
$F_{bias}<0$ corresponds to a solar-like FIP effect, and $F_{bias}>0$
corresponds to an inverse FIP effect.  It is important to realize that
the coronal abundances quantified by $F_{bias}$ are relative
abundances, not absolute ones.  Computing absolute abundances, i.e.,
relative to H, requires an additional analysis of line-to-continuum
ratio, which represents a substantial source of systematic error.
Sticking with relative abundances allows us to avoid inserting this
uncertainty into the $F_{bias}$ quantity, but it does mean that, for
example, $F_{bias}<0$ does not tell you whether the low-FIP elements
are being enhanced in the corona (as in the case of the Sun) or if the
high-FIP elements are being depleted.

     Figure~3 shows that for the WL10 sample of stars there is a
strong correlation between $F_{bias}$ and Fe~XVII/O~VIII.  This is
quantified by a second order polynomial fit, where if $x\equiv
\log{({\mbox{\rm Fe~XVII/O~VIII}})}$, $F_{bias}=0.344-1.926x+0.994x^2$.
Thus, for GJ~338A and GJ~338B, we can measure Fe~XVII/O~VIII ratios from
Table~1 and then infer $F_{bias}$ from the polynomial fit in
Figure~3.  Error boxes in Figure~3 show the results independently
for GJ~338A and GJ~338B.  However, given that Fe~XVII/O~VIII is so
similar for the two stars, and given that they have the same spectral
type, we deem it worthwhile to simply sum the line fluxes of the two
stars and compute a single Fe~XVII/O~VIII ratio and $F_{bias}$ value
for GJ~338AB.  The result, $x=0.020\pm 0.130$ and
$F_{bias}=0.31^{+0.26}_{-0.23}$, is shown in the figure as well.
For GJ~338AB we find that $F_{bias}>0$, indicating an inverse FIP
effect.  If stellar activity is quantified by $F_X$, the GJ~338 stars
are now easily the least active stars known to have an inverse FIP
effect present in their coronae.

     There is an important caveat about the correlation between
$F_{bias}$ and Fe~XVII/O~VIII in Figure~3: the relation is inferred
from a particular sample of moderately active stars with similar
emission measure distributions.  Stars with coronal temperatures much
different than these may not be consistent with this relation.  In
order for this relation to be applicable to GJ~338AB, we need to have
reason to believe that GJ~338AB's coronal temperatures are comparable
to those of the WL10 sample of stars.  The best diagnostic of coronal
temperature available to us for GJ~338AB is the flux ratio of the
O~VIII and O~VII lines in Table~1.  Uncertainties in the individual
line measurements are very high, but if the O~VIII line fluxes for
both stars are summed, and then divided by the sum of all O~VII lines
from both stars, the result is a flux ratio of O~VIII/O~VII=0.8.
These ratios are near the lower
bound observed within the WL10 sample.  They are still comparable,
though, being particularly close to the ratio seen for 36~Oph~B, for
example \citep[see][]{bew06}, suggesting similar coronal temperatures.
Thus, we think the $F_{bias}$ measurement for GJ~338AB provided by
Figure~3 should be good to within the broad error bars.  In contrast,
we note in passing that line measurements from LETGS spectra of
$\alpha$~Cen~A and B (G2~V+K1~V) by \citet{ajjr03} imply that these
inactive stars have much lower O~VIII/O~VII ratios, indicative of much
cooler coronae, and as a consequence their $F_{bias}$ values are {\em
not} consistent with the relation in Figure~3 (see Table~2 and
discussion below), demonstrating that not all stars will follow this
relation.

     In order to put the new measurement into context, we have in
Table~2 compiled a list of $F_{bias}$ quantities that can be computed
from past coronal abundance analyses of X-ray spectra, keeping in mind
that we are interested in main sequence stars with $\log L_X<29$.
This mostly consists of the WL10 sample of stars, but as that paper
did not provide a tabulated list of stars and $F_{bias}$ numbers, we
take the opportunity to do so here.  To the WL10 sample, we add
$\alpha$~Cen AB (G2~V+K1~V) from \citet{ajjr03}, and our new GJ~338AB
result.  Finally, at the bottom of the table, we list a few
representative very active stars with $\log L_X>29$, which we will use
to illustrate how the coronal abundances of such stars behave
differently from the less active main sequence stars.

     The fourth column of Table~2 lists radii for our sample of stars.
For most of the G and K dwarfs, the radii are computed using the
Barnes-Evans relation \citep{tgb78}, except for $\alpha$~Cen~A and B,
for which direct measurements of radii are available \citep{pk03}.
With the exception of GJ~338AB and Proxima~Cen, whose radii have
already been discussed earlier, we use \citet{jm08} as the source for
the stellar radii of the M dwarfs.  The X-ray luminosities in column 5
of Table~2 are for the ROSAT PSPC bandpass of $0.1-2.4$~keV, and most
are from ROSAT PSPC measurements made during the ROSAT All-Sky Survey
\citep{js04}.  For the Sun, we assume $\log L_X=27.35$, the middle of
the range computed by \citet{pgj03}.  \citet{tra09} estimate that
$\alpha$~Cen~A and B have average X-ray fluxes about a factor of two
below and above that of the Sun, respectively, so that is what is
assumed for $\alpha$~Cen~AB in Table~2.  The sixth column lists X-ray
surface fluxes computed from the X-ray luminosities and stellar radii
listed in columns four and five.

     The last column of Table~2 lists the references for the coronal
abundance analyses for these stars.  There are two different
quantifications of coronal FIP bias listed in the table.  The first is
a simple coronal Ne/Fe abundance ratio, with no attempt to normalize
this to any assumed photospheric abundances.  And finally, of course,
the more complex $F_{bias}$ quantity that we have described in detail
above.  Although we use coronal abundances measured by others, in
computing $F_{bias}$ we
are careful to apply our own self-consistent assumptions about assumed
reference photospheric abundances, as described above.  \citet{cap04}
is the source for all the photospheric abundances assumed for the G
and K stars, except for EK~Dra and AB~Dor, for which we simply rely on
the same photospheric abundances assumed in the original coronal
abundance analysis.

     Figure~4a reproduces the FBST relation
from WL10, but with the addition of extra data points.  These
include our new GJ~338AB data point, which is nicely consistent
with the relation.  The addition of $\alpha$~Cen~A and B
appears to significantly increase the scatter, suggesting
that $\alpha$~Cen~AB may be somewhat
discrepant.  It is possible that this is because these stars are
significantly less active and have significantly cooler coronae than
any of the other stars in the sample, except for the Sun.  Finally,
the three representative very active stars listed in Table~2 are
also plotted in Figure~4a, demonstrating that stars with $\log L_X>29$
lie above the FBST relation defined by the less active stars.

     Figure~4b is analogous to Figure~4a, but uses the simpler FIP
bias indicator, the coronal Ne/Fe ratio.  One purpose of this figure is
to demonstrate that even with a simpler FIP bias quantifier, with no
attempt whatsoever to normalize to photospheric abundances, you can
still see the same correlation in Figure~4b as in 4a, demonstrating that
the FBST correlation is not a product of assumptions about
photospheric abundances.  \citet{jsf09} hypothesized that inverse FIP
effects reported in the literature might be an artifact of assuming
solar photospheric abundances for these stars, as none of these stars
have measured photospheric abundances.  However, the consistency the M
dwarfs show with the overall spectral type dependence in Figure~4
would argue against this interpretation, as do observations of flares
in such stars \citep[e.g., EV Lac;][]{jml09a}, where composition
changes during the event, interpreted as the result of chromospheric
evaporation, suggest a photospheric composition similar to that of the
Sun.  Figure~4b also allows us to compare our main sequence Ne/Fe
ratios with those measured for T~Tauri stars by \citet{mg07}.
\citet{mg07} reported a spectral type dependence of coronal abundance
for T~Tauri stars, which parallels the main sequence FBST relation,
basically where the very active main sequence stars lie in the
figure.

     Figures~4c and 4d demonstrate explicitly that in our sample of
stars there is no correlation of $F_{bias}$ with activity at all,
regardless of whether $\log L_X$ or $\log F_X$ is used as the activity
quantifier.  By itself, Figure~4d might suggest that the least active
stars tend to have low $F_{bias}$ values, but this is a selection
effect.  We would argue on the basis of Figure~4a that the upper left
corner of Figure~4d is actually full of inactive M dwarfs that are
simply too faint to observe spectroscopically in X-rays.  The GJ~338AB
observation presented here basically represents an attempt to push
farther into this corner of Figure~4d.  Replacing $\log F_X$ with
$\log L_X/L_{bol}$, which is another commonly used activity
diagnostic, yields results qualitatively similar to Figure~4d, but
with the M dwarfs shifted further to the right relative to the other
stars.

     We have been quoting $\log L_X=29$ as the boundary between stars
that obey the FBST relation and stars that do not, but intuitively one
would think that such a border would be better defined in terms of a
normalized activity diagnostic like $\log F_X$ or $\log L_X/L_{bol}$,
considering that the sample of stars contains stars of various sizes.
However, in Figure~4d it is not easy to clearly draw a vertical line
separating the filled symbols (stars that are consistent with the FBST
relation) and the open symbols (stars not consistent with the FBST
relation).  Note that the situation is even worse if $\log L_X/L_{bol}$
is plotted instead of $\log F_X$, with the M dwarfs even further to
the right relative to the other data points.
One could perhaps draw the line at $\log F_X=7.0$, which
would only place EV~Lac on the wrong side of the line.  Nevertheless,
the $\log L_X=29.1$ boundary shown in Figure~4c seems to work better,
so we will in the future quote $\log L_X=29.1$ as the activity
threshold where activity dependence of coronal abundances starts to
become apparent.  (With several stars in our sample right at $\log
L_X=29.0$, it seems wise to conservatively move the divider to $\log
L_X=29.1$.)

\section{MODELING THE INVERSE FIP EFFECT}

     One of the striking features of the FBST relation illustrated in
Figure 4 is how the coronal abundance anomaly smoothly changes from
solar-like FIP bias at spectral types G to early K, to inverse FIP in
M dwarfs.  This suggests that a model for solar-like FIP fractionation
should, with suitably chosen parameters, be capable of predicting an
inverse FIP effect.  \citet{jml04} reviewed the various solar FIP
models available at that time, and argued that only the model based on
the ponderomotive force arising from MHD waves could also plausibly
explain the inverse FIP effect.  Here we elaborate on this suggestion
and provide a semi-quantitative illustration of how such a model would
work.  We defer a fuller exposition, designed to match specific stars,
to later papers.

     The model described by \citet{jml04,jml09} and most recently by
\citet{jml12} for the FIP effect assumes that Alfv\'en waves of
amplitude approximately 50 km s$^{-1}$ are generated in a coronal loop
at the resonant frequency, and remain trapped in the loop ``resonant
cavity''.  Upon reflection from chromospheric footpoints, the waves
develop a ponderomotive force in the steep density gradients there,
and this force, acting on chromospheric ions (but not neutrals),
preferentially accelerates these ions up into the
corona. \citet{jml12} studies the fractionations produced by waves on
and off resonance, and \citet{rakowski12} extend this to different
loop lengths and magnetic fields, concentrating mainly on the
fractionation of He with respect to O.  These works only consider
coronal Alfv\'en waves, with chromospheric acoustic waves included in
an ad hoc manner.  When these upcoming chromospheric acoustic waves
are allowed to mode convert, at the layer where sound and Alfv\'en
speeds are equal, to what in the magnetically dominated upper
chromospheric become fast mode waves, inverse FIP fractionation can
result.  This arises because the fast mode waves undergo reflection
back downwards as the Alfv\'en speed increases, giving rise to a
downwards directed ponderomotive force than can compete with that due
to the coronal Alfv\'en waves.

     We treat the fast mode waves as approximately isotropic in the
upward moving hemisphere.  Then the fraction reflected at
chromospheric height $z$ is
\begin{equation}
f_R\left(z\right)\simeq\sqrt{1-{c_s^2\left(z_{\beta =1}\right)\over
  V_A^2\left(z\right)+c_s^2\left(z\right)}}
\end{equation}
where $z_{\beta =1}$ is the chromospheric height where mode conversion
occurs.  The ponderomotive acceleration due to fast mode waves is
then
\begin{equation}
a={c^2\over 4}{\partial\over\partial z}\left(\delta E^2\over
B^2\right)={\delta v^2\over 2}\left(1-f_R\right){1\over\delta
v}{\partial\delta v\over\partial z}-{\delta v^2\over 4} {\partial
f_R\over\partial z},
\end{equation}
where $\delta v$ is the wave amplitude in km~s$^{-1}$.
The two terms represent an upwards contribution arising as the fast
mode waves increase in amplitude as they propagate through lower
density plasma, and the downwards contribution arising from fast mode
wave reflection.  Evaluating
\begin{equation}
{\partial f_R\over\partial z} ={c_s^2\left(z_{\beta =1}
  \right)V_A\over\left(V_A^2+c_s^2\right)^2f_R}{\partial
  V_A\over\partial z}= {c_s^2\left(z_{\beta
  =1}\right)V_A^2\over\left(V_A^2+c_s^2\right)^2f_R} \left({1\over
  H_B}-{1\over 2H_D}\right)
\end{equation}
and assuming from the WKB approximation
\begin{equation}
{1\over \delta v}{\partial \delta v\over\partial z}={-1\over
  2H_B}-{1\over 4H_D},
\end{equation}
where $H_D$ and $H_B$ are the signed density and magnetic field scale
heights, we find
\begin{equation}
a={\delta v^2\over f_R}\left\{\left(f_R-1\right)\left(-{1\over8H_D}-
  {1\over 4H_B}\right) +{c_s^2\left(z_{\beta =1}
  \right)\over\left(V_A^2+c_s^2\right)^2}\left(-{c_s^2\over 8H_D}-
  {c_s^2\over 4H_B}-{V_A^2\over 2H_B}\right)\right\}.
\end{equation}
Remembering that both $H_D$ and $H_B$ are negative, and $f_R < 1$, the
first term in curly brackets is negative, giving rise to an inverse
FIP effect, and the second term is positive, giving the more usual FIP
effect.  In conditions where $V_A >> c_s$, we expand $f_R$ as a Taylor
series in the small quantity $c_s^2/V_A^2$, and to leading order
in this quantity find that an overall downwards
pointed ponderomotive acceleration requires $\left|H_D\right| <
\left|H_B\right|/6$ in this simple model.

     Additional reflection of fast mode waves from, e.g., density
fluctuations (not included in this model) would relax the requirement.
So too would fast mode waves spreading out laterally from a
horizontally localized source.  Even so, equation (5) implies that
inverse FIP effect is more likely to be found in stars with minimal
magnetic field expansion through the chromosphere, consistent with
observations of active M dwarfs.  While the magnetic fields measured
in such stars are similar to those in the Sun, the filling factor is
higher \citep[e.g.,][]{donati09}, allowing less volume for expansion
with increasing altitude.  However, it is less certain whether this
argument applies to less active M dwarfs like GJ~338AB.

     These fast mode wave are assumed to derive from mode conversion
of $p$-modes at this layer.  The acoustic waves have an energy
transmission coefficient of \citep{cally08}
\begin{equation}
T=\exp\left(-\pi \left|{\bf k}\right|\sec\theta\sin ^2\theta\over
  \left[d\left(V_A^2/c_s^2\right)/dz\right]_{\beta =1}\right)
\end{equation}
in vertical magnetic field, where $\theta$ is the polar angle of the
wavevector.  Similarly to the Alfv\'en waves, we model the acoustic
wave energy density $U_{ac}$ associated with the fast mode wave energy
density $U_{fm}$ by $U_{ac}\left(z\right)=6U_{fm}\left(z_{\beta
=1}\right)c_s\left(z\right)^2 /V_A\left(z\right)^2$.  The factor of 6
is chosen to match the acoustic and Alfv\'en or fast mode amplitudes
given in \citet{cranmer07} and \citet{khomenko11}, and is consistent
with equation (6) for
$\left|k\right|/\left[d\left(V_A^2/c_s^2\right)/dz\right]_{\beta
=1}=0.145$ when $\theta = 32^{\circ}$.  The last factor of
$c_s\left(z\right)^2 /V_A\left(z\right)^2$ approximates the continuing
reflection of acoustic waves by the increasing cut off frequency above
the mode conversion layer \citep[see, e.g.,][]{barnes01}.

     Figure 5 illustrates a calculation designed to give an inverse
FIP effect.  The figure is analogous to Figure~3 of \citet{jml12},
which illustrates a model of a solar-like FIP effect.  A loop of
length 100,000~km, with a 100~G magnetic field is considered. The
magnetic field is uniform through the chromosphere, and a fast mode
wave amplitude of 20 km s$^{-1}$ is included at the $\beta = 1$ layer,
which is allowed to propagate and refract as described above.  The
model chromosphere derives from the solar model of \citet{avrett08}.
Future work should implement a model {\em stellar} chromosphere.  The
top left panel shows the variation of the perturbations $\delta v$ and
$\delta B/\sqrt{4\pi\rho}$ associated with the coronal Alfv\'en wave,
and the bottom left panel shows the Alfv\'en wave energy fluxes, both
upward and downwards directed.  The dotted line in the bottom right
panel shows the difference in wave energy fluxes, and should be
horizontal in the absence of wave damping or growth if energy is
conserved.  The top right panel shows the ponderomotive acceleration.
The positive contribution in the upper chromosphere (solid curve)
comes from the coronal Alfv\'en waves.  The negative contribution
lower down (dashed curve) comes from the total internal reflection of
fast mode waves.  The dotted curve gives the amplitude of acoustic
waves through the chromosphere, modeled as outlined above.  The bottom
right panel gives the FIP fractionations for the ratios S/O, He/O,
Mg/O and Fe/O.  The He/O ratio remains unchanged, but all others
display an inverse FIP effect.

     The transition from FIP effect to inverse FIP effect with
increasing fast mode wave amplitude at the $\beta =1$ layer is
illustrated in Table 3.  In the absence of fast mode waves, a small FIP
effect results, with the inverse FIP effect becoming gradually more
dominant as amplitude increases.

\section{SUMMARY}

     We have analyzed {\em Chandra} observations of the GJ~338AB
binary, representing the least active M dwarfs ever observed
spectroscopically in X-rays.  Our results are summarized as follows:
\begin{enumerate}
\item Despite the limited number of lines detected in the spectra,
  we demonstrate that the coronae of the two GJ~338 stars exhibit an
  inverse FIP effect, making these the least active stellar coronae
  known to show this phenomenon.
\item The inverse FIP effect observed for GJ~338AB is consistent with
  expectations from the FBST relation of WL10, which suggests that
  {\em all} M dwarfs, both active and inactive, should exhibit an inverse
  FIP effect.
\item A compilation of coronal abundance measurements for main sequence
  stars (beyond that of WL10) is used to demonstrate that these stars
  exhibit no significant correlation between coronal abundance and
  activity, regardless of whether $\log L_X$ or $\log F_X$ is used as
  the activity diagnostic.
\item Past work has demonstrated that MHD waves propagating through
  coronal loops can generate a ponderomotive force capable of
  fractionating elements in a manner consistent with the solar-like
  FIP effect \citep{jml04,jml09,jml12}.  Building on this theoretical
  framework, we have shown that such models can also yield an inverse
  FIP effect in certain circumstances, depending on where MHD waves in
  the stellar atmosphere are generated, and on where they are
  transmitted and reflected.  In particular, waves propagating upward
  from the chromosphere and reflecting back down rather than down from
  the corona and reflecting back up tend to lead to inverse FIP.  In
  the future, we will develop models specifically tailored for M
  dwarfs such as GJ~338AB, in order to explore in more detail what
  characteristics of M dwarf atmospheres lead their coronae to have an
  inverse FIP effect instead of a solar-like FIP effect.
\end{enumerate}

\acknowledgments

We would like to thank Brad Wargelin and Vinay Kashyap for their
assistance in identifying the source of the asymmetry in the GJ~338A
image.  Support for this work was provided by NASA through ATP award
NNH11AQ23I and Chandra Award Number GO1-12012Z issued by the Chandra
X-ray Center (CXC).  MK is a member of CXC, which is operated by the
Smithsonian Astrophysical Observatory for and on behalf of NASA under
contract NAS8-03060.

\clearpage

\begin{deluxetable}{lccccc}
\tabletypesize{\scriptsize}
\tablecaption{{\em Chandra} Line Measurements}
\tablecolumns{6}
\tablewidth{0pt}
\tablehead{
  \colhead{Ion} & \colhead{$\lambda_{obs}$} & \multicolumn{2}{c}{Counts} &
    \multicolumn{2}{c}{Flux (10$^{-5}$ ph cm$^{-2}$ s$^{-1}$)} \\
  \colhead{} & \colhead{(\AA)} & \colhead{GJ 338A} & \colhead{GJ 338B} &
    \colhead{GJ 338A} & \colhead{GJ 338B}}
\startdata
Ne IX  &13.6& $75.8\pm 27.1$ & $48.8\pm 27.1$ &$2.81\pm1.00$ & $1.81\pm1.00$ \\
Fe XVII&15.1& $34.3\pm 22.9$ & $40.2\pm 26.7$ &$1.25\pm0.84$ & $1.47\pm0.98$ \\
Fe XVII&17.0& $58.6\pm 24.1$ & $61.6\pm 21.5$ &$2.47\pm1.02$ & $2.60\pm0.91$ \\
O VIII &19.0& $81.3\pm 21.7$ & $89.6\pm 20.6$ &$3.44\pm0.92$ & $3.77\pm0.87$ \\
O VII  &21.6& $50.4\pm 22.4$ & $29.7\pm 21.7$ &$3.29\pm1.46$ & $1.92\pm1.40$ \\
O VII  &22.1& $33.1\pm 18.7$ & $20.8\pm 18.3$ &$2.20\pm1.24$ & $1.38\pm1.21$ \\
N VII  &24.8& $35.2\pm 20.1$ & $34.4\pm 21.5$ &$2.39\pm1.36$ & $2.33\pm1.46$ \\
C VI   &33.7& $41.8\pm 20.4$ & $22.8\pm 14.3$ &$3.79\pm1.85$ & $2.06\pm1.29$ \\
\enddata
\end{deluxetable}

\begin{deluxetable}{lcccccccc}
\tabletypesize{\scriptsize}
\tablecaption{List of FIP Bias Assessments for Main Sequence Stars}
\tablecolumns{9}
\tablewidth{0pt}
\tablehead{
  \colhead{Star} & \colhead{Alternate} & \colhead{Spectral} &
    \colhead{Radius}        & \colhead{$\log L_X$}      &
    \colhead{$\log F_X$}               & \colhead{Ne/Fe} &
    \colhead{$F_{bias}$} & \colhead{Ref.} \\
  \colhead{}     & \colhead{Name}      & \colhead{Type}     &
    \colhead{(R$_{\odot}$)} & \colhead{(ergs s$^{-1}$)} &
    \colhead{(ergs cm$^{-2}$ s$^{-1}$)} & \colhead{}      &
    \colhead{}            & \colhead{}}
\startdata
$\beta$ Com &HD 114710 & G0 V    & 1.08 & 28.21 & 5.36 & 0.73 & $-0.668$ & 1 \\
$\pi^1$ UMa &HD 72905  & G1 V    & 0.91 & 28.97 & 6.27 & 1.60 & $-0.645$ & 1 \\
$\chi^1$ Ori&HD 39587  & G1 V    & 0.98 & 28.99 & 6.22 & 1.91 & $-0.555$ & 1 \\
Sun         & ...      & G2 V    & 1.00 & 27.35 & 4.57 & 1.02 & $-0.600$ & 2 \\
$\alpha$ Cen A&HD 128620&G2 V    & 1.22 & 27.00 & 4.04 & 1.41 & $-0.410$ & 3 \\
$\kappa$ Ceti&HD 20630 & G5 V    & 0.98 & 28.79 & 6.02 & 2.51 & $-0.462$ & 1 \\
$\xi$ Boo A &HD 131156A& G8 V    & 0.83 & 28.86 & 6.24 & 3.80 & $-0.344$ & 4 \\
70 Oph A    &HD 165341A& K0 V    & 0.85 & 28.27 & 5.63 & 2.40 & $-0.403$ & 5 \\
36 Oph A    &HD 155886 & K1 V    & 0.69 & 28.10 & 5.64 & 4.90 & $-0.250$ & 5 \\
36 Oph B    &HD 155885 & K1 V    & 0.59 & 27.96 & 5.63 & 3.80 & $-0.328$ & 5 \\
$\alpha$ Cen B&HD 128621&K1 V    & 0.86 & 27.60 & 4.95 & 1.44 & $-0.478$ & 3 \\
$\epsilon$ Eri&HD 22049& K2 V    & 0.78 & 28.32 & 5.75 & 4.68 & $-0.050$ & 5 \\
$\xi$ Boo B &HD 131156B& K4 V    & 0.61 & 27.97 & 5.62 & 7.41 & $-0.185$ & 4 \\
70 Oph B    &HD 165341B& K5 V    & 0.66 & 28.09 & 5.67 & 8.71 & $0.138$  & 5 \\
GJ 338AB    & ... &M0 V+M0 V& 0.59+0.58 & 27.92 & 5.30 & ...  & $0.305$  & 6 \\
EQ Peg A    & GJ 896A  & M3.5 V  & 0.35 & 28.71 & 6.84 &16.12 & $0.450$  & 7 \\
EV Lac      & GJ 873   & M3.5 V  & 0.30 & 28.99 & 7.25 &13.92 & $0.474$  & 7 \\
EQ Peg B    & GJ 896B  & M4.5 V  & 0.25 & 27.89 & 6.31 &14.67 & $0.417$  & 7 \\
AD Leo      & GJ 388   & M4.5 V  & 0.38 & 28.80 & 6.86 &18.25 & $0.536$  & 7 \\
Proxima Cen & GJ 551   & M5.5 V  & 0.15 & 27.22 & 6.08 &11.98 & $0.471$  & 7 \\
\multicolumn{9}{l}{\underline{Very Active Star Sample ($\log L_X>29$)}} \\
EK Dra      &HD 129333 & G1.5 V  & 0.89 & 29.93 & 7.25 & 3.71 & $-0.277$ & 1 \\
AB Dor      &HD 36705  & K0 V    & 0.79 & 30.06 & 7.48 &17.11 & $0.488$  & 8 \\
AU Mic      &HD 197481 & M1 V    & 0.61 & 29.62 & 7.27 &24.30 & $0.695$  & 7 \\
\enddata
\tablerefs{(1) Telleschi et al.\ 2005. (2) Feldman \& Laming 2000. (3)
  Raassen et al.\ 2003. (4) Wood \& Linsky 2010 (WL10). (5)
  Wood \& Linsky 2006. (6) This paper. (7) Liefke et al.\ 2008. (8)
  G\"{u}del et al.\ 2001.}
\end{deluxetable}

\begin{deluxetable}{lcccc}
\tabletypesize{\scriptsize}
\tablecaption{Modeled Coronal Abundance Ratios}
\tablecolumns{5}
\tablewidth{0pt}
\tablehead{
  \colhead{Ratio} &  \multicolumn{4}{c}{Fast Mode Wave Amplitude} \\
  \colhead{} & \colhead{0 km s$^{-1}$} &\colhead{5 km s$^{-1}$} &
    \colhead{10 km s$^{-1}$} &\colhead{20 km s$^{-1}$} }
\startdata
He/O& 0.98& 0.98& 0.98& 0.98\\
C/O & 1.00& 1.00& 1.00& 1.00\\
N/O & 1.00& 1.00& 1.00& 1.00\\
Ne/O& 1.00& 1.00& 1.00& 1.00\\
Mg/O& 1.04& 0.90& 0.80& 0.70\\
Si/O& 1.04& 0.89& 0.80& 0.71\\
Ar/O& 1.00& 1.00& 1.00& 1.00\\
Ca/O& 1.06& 0.89& 0.79& 0.68\\
Fe/O& 1.07& 0.88& 0.78& 0.67\\
\enddata
\end{deluxetable}

\clearpage

\begin{figure}[t]
\plotfiddle{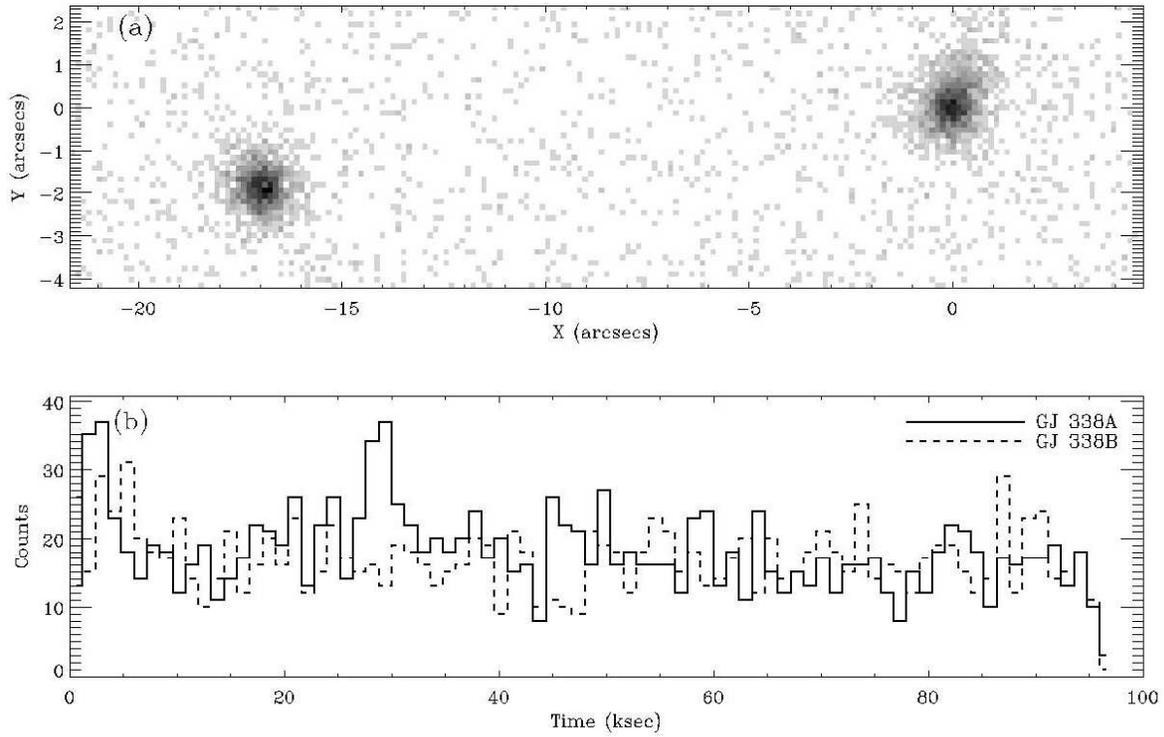}{-3.0in}{0}{350}{280}{10}{-50}
\caption{(a) The zeroth-order image of the GJ~338 binary
  from the {\em Chandra} LETGS observation, with GJ~338A being the
  source on the right.  North is up in the figure.  An excess of
  emission seen northwest of GJ~338A is due to an instrumental
  artifact.  (b) X-ray light curves measured for GJ~338A and
  GJ~338B from the zeroth-order image, using 20 minute time bins.}
\end{figure}


\clearpage

\begin{figure}[t]
\plotone{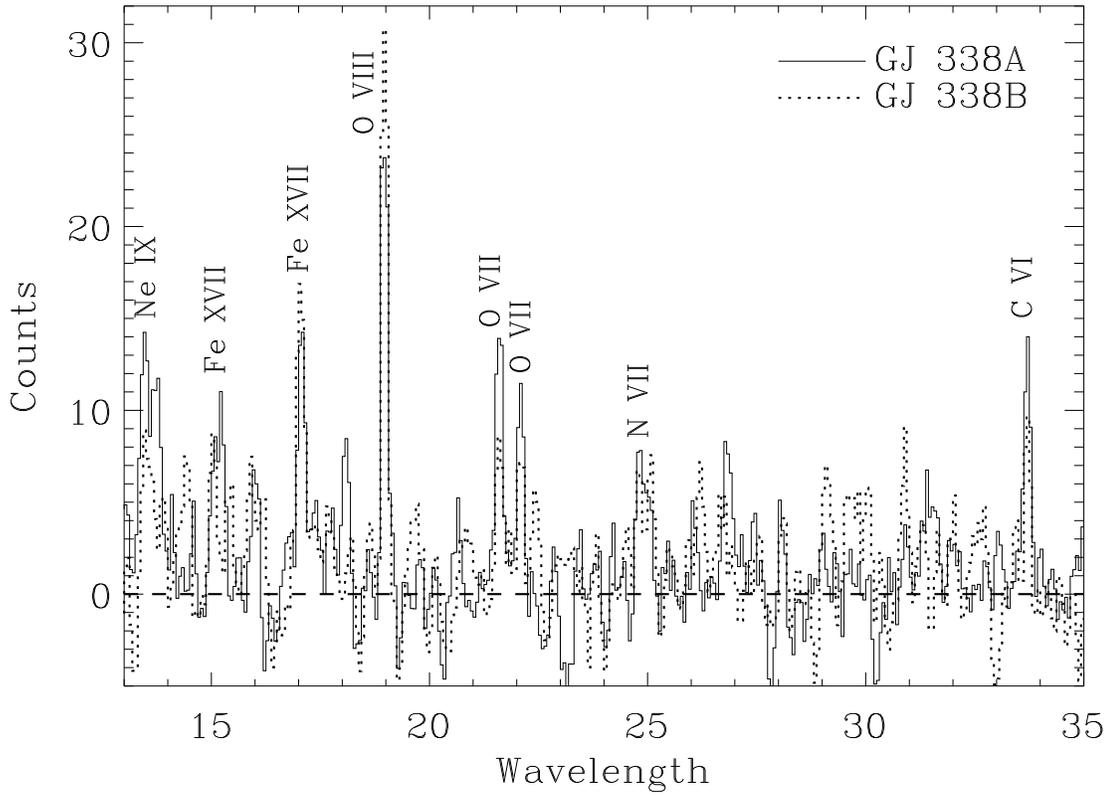}
\caption{{\em Chandra} LETGS X-ray spectra of GJ~338A and GJ~338B.
  The wavelength is in {\AA}ngstroms.}
\end{figure}

\clearpage


\begin{figure}[t]
\plotone{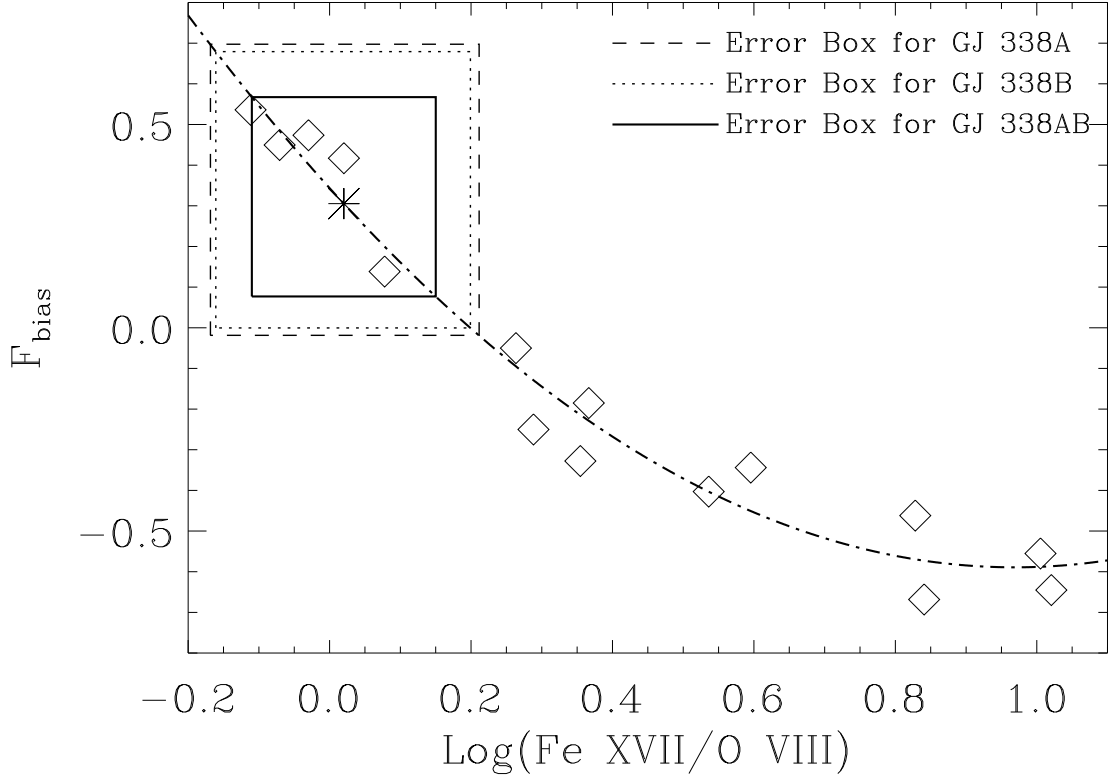}
\caption{The coronal FIP bias, $F_{bias}$, is plotted versus the
  logarithmic flux ratio of Fe~XVII 15--17~\AA\ lines to the O~VIII 19.0~\AA\
  line, for the sample of moderately active main sequence stars
  studied by WL10.  Values of $F_{bias}<0$ correspond to a solar-like
  FIP effect, while $F_{bias}>0$ corresponds to an inverse FIP effect.
  A second-order polynomial has been fitted to these data points
  (dot-dashed line).  Combining this relation with the Fe~XVII/O~VIII
  ratios measured for GJ~338 yields the displayed error boxes, which
  are shown for the binary as a whole, and for each member
  separately.}
\end{figure}

\clearpage

\begin{figure}[t]
\plotone{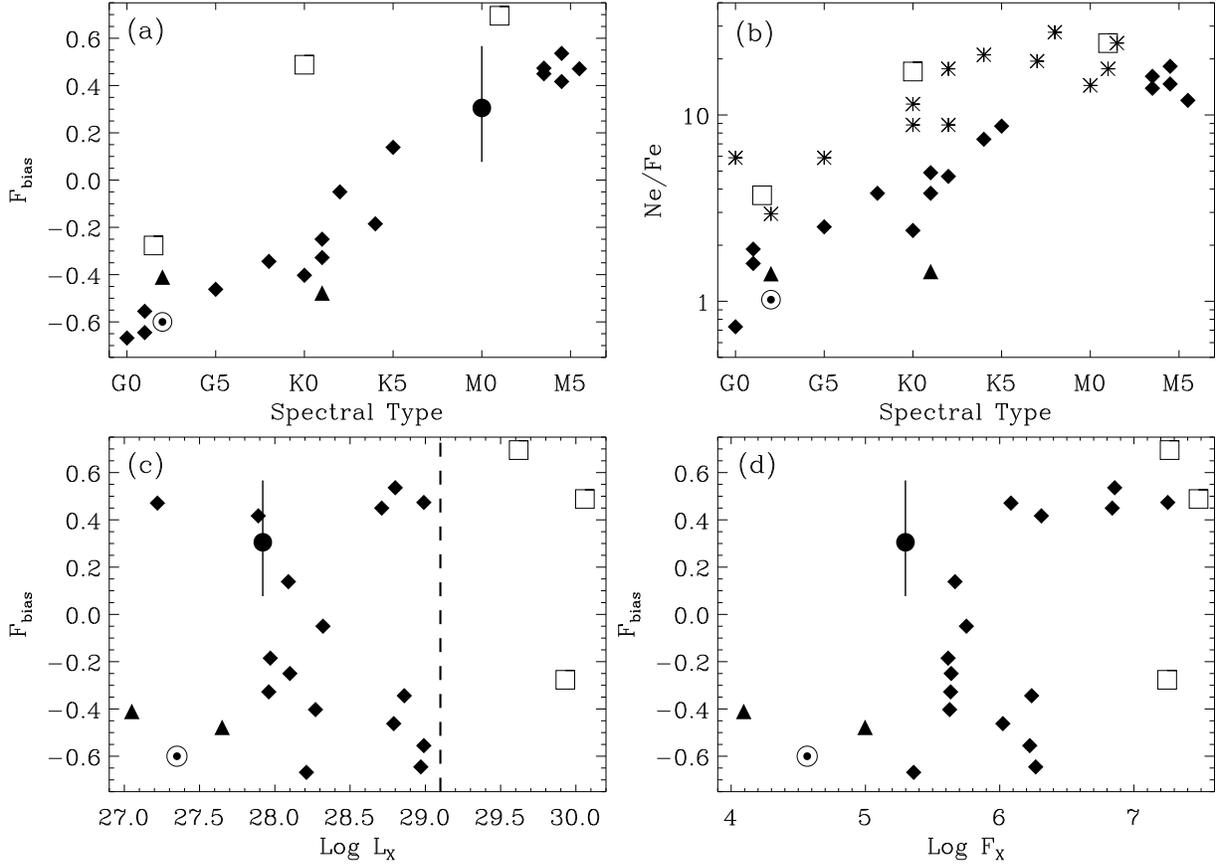}
\caption{(a) A plot of $F_{bias}$ versus spectral type for the WL10
  sample of main sequence stars with $\log L_X<29$ (diamonds),
  combined with measurements of $\alpha$~Cen A and B (triangles) and
  our new measurement for GJ~338AB (circle).  See Table~2 for a list
  of the stars.  A strong correlation of $F_{bias}$ with spectral type
  is apparent, i.e., the FBST relation.  The open squares are examples
  of three main sequence stars above the $\log L_X=29$ threshold, all
  of which lie above the FBST relation seen for the less active stars.
  (b) The coronal Ne/Fe ratio is plotted versus spectral type for both
  the Table~2 sample of main sequence stars, and for T~Tauri stars
  from G\"{u}del et al.\ (2007) (asterisks).  (c) Plot of $F_{bias}$
  versus X-ray luminosity. A vertical line at $\log L_X=29.1$
  separates the stars that follow the FBST relation in (a) and those
  that do not.  (d) Plot of $F_{bias}$ versus X-ray surface flux.}
\end{figure}

\begin{figure}[t]
\plotone{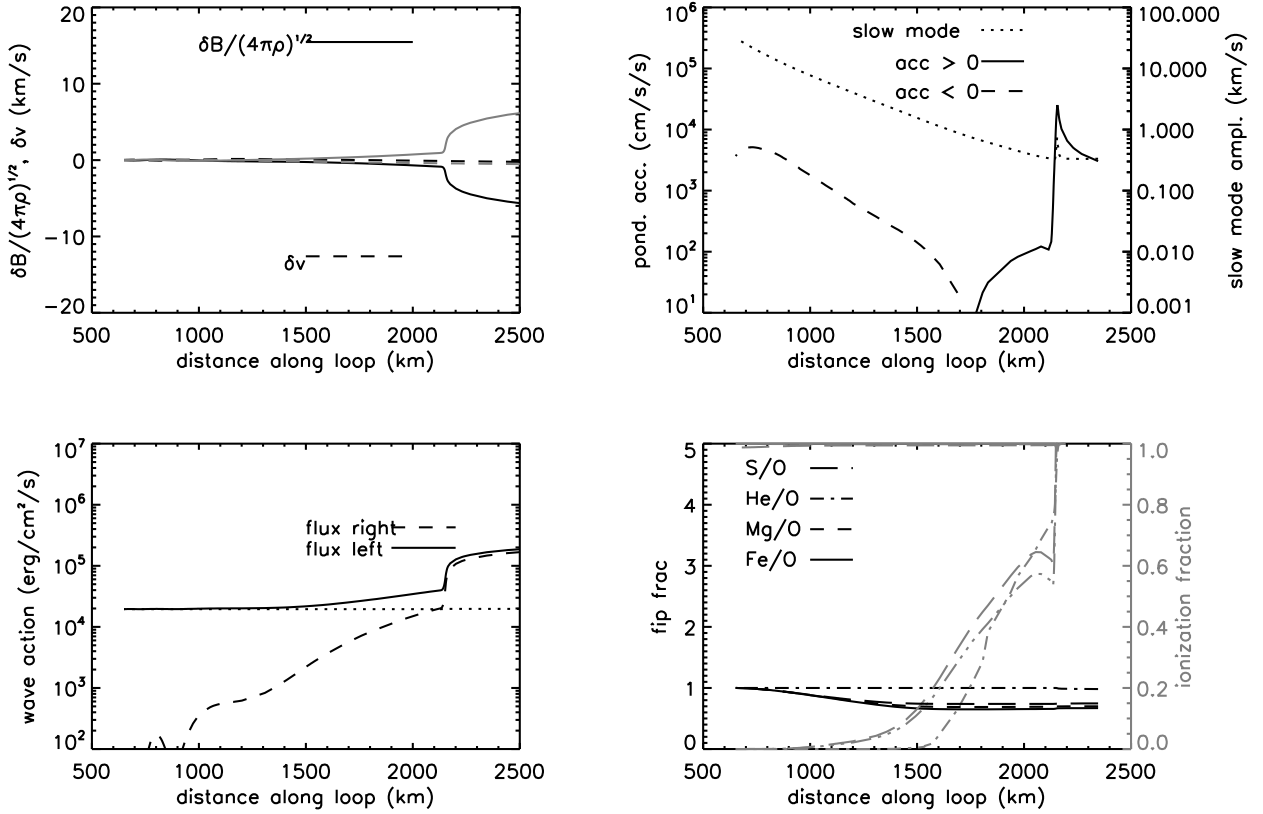}
\caption{Illustration of a model for the inverse FIP effect, which can
  be compared with Figure~3 from Laming (2012) illustrating a model
  for a solar-like FIP effect.  The top left panel shows the variables
  $\delta v$ and $\delta B/\sqrt{4\pi\rho}$ for the coronal Alfv\'en
  wave, with black and gray lines corresponding to real and imaginary
  parts.  The bottom left panel shows the upgoing (dashed curve) and
  downgoing (solid curve) wave energy fluxes.  The dotted line shows
  their difference, and should be horizontal in the absence of wave
  growth or damping, if energy is conserved.  The top right panel
  shows the ponderomotive acceleration.  The positive contribution in
  the upper chromosphere (solid curve) comes from the coronal Alfv\'en
  waves.  The negative contribution lower down (dashed curve) comes
  from the total internal reflection of fast mode waves.  The dotted
  curve gives the amplitude of acoustic waves through the
  chromosphere.  The gray lines in the bottom right panel show
  ionization fractions for S, He, Mg, and Fe, with the extra
  triple-dot-dashed line showing O as well.  The Mg and Fe ionization
  fractions are both near 1 throughout the loop.  Most importantly,
  the solid lines show various abundance ratios: He/O remains
  unchanged, but all others indicate an inverse FIP effect.}
\end{figure}

\end{document}